\documentclass{aa}
\newif\ifpdf\ifx\pdfoutput\undefined\pdffalse\else\pdfoutput=1\pdftrue\fi
\ifpdf
  \usepackage[pdftex]{graphicx}
\else
  \usepackage{graphicx}
\fi
\usepackage{natbib}
\usepackage{amsmath,amssymb}
\usepackage{mathrsfs}
\usepackage{txfonts}

\bibpunct{(}{)}{;}{a}{}{,} 

\def\mdrad{\ensuremath{\langle r_{\mathrm{d}}\rangle}}
\def\CtHt{\ensuremath{\mbox{C}_2\mbox{H}_2}}
\def\CtH{\ensuremath{\mbox{C}_2\mbox{H}}}
\def\Ct{\ensuremath{\mbox{C}_2}}
\def\rhod{\ensuremath{\rho_{\mathrm{d}}}}

\def\teg{\ensuremath{T_{\mathrm{g}}}}

\def\vtp{\ensuremath{\tilde{v}}}
\def\wtp{\ensuremath{\tilde{w}}}
\def\vtpr{\ensuremath{\tilde{v}_{i(,m)}}}
\def\wtpr{\ensuremath{\tilde{w}_{i(,m)}}}

\def\kms{\ensuremath{\,\mbox{km}\,\mbox{s}^{-1}}}

\def\CtoO{\ensuremath{\varepsilon_{\mathrm{C}}/\varepsilon_{\mathrm{O}}}}
\def\drhog{\ensuremath{\rho_{\mathrm{d}}/\rho}}

\def\fcond{\ensuremath{f_{\mathrm{cond}}}}
\def\vdri{\ensuremath{v_{\mathrm{D}}}}
\def\vdria{\ensuremath{v_{\mathrm{D}}(a)}}

\def\teff{\ensuremath{T_{\mathrm{eff}}}}
\def\deltaup{\ensuremath{\Delta\,u_{\mathrm{p}}}}

\def\sistd{\ensuremath{\sigma_{\mathrm{s}}}}
\def\mmdot{\ensuremath{\langle\dot{M}\rangle}}
\def\muinf{\ensuremath{\langle u_{\infty}\rangle}}
\def\mfcond{\ensuremath{\langle f_{\mathrm{cond}}\rangle}}
\def\mdrhog{\ensuremath{\langle\drhog\rangle}}

\def\mvdri{\ensuremath{\langle v_{\mathrm{D}}\rangle}}

\newcommand{\rSaHoa}{Paper~I}
\newcommand{\rSaHob}{Paper~II}
\newcommand{\rKrSe}{KS97}
\newcommand{\rGaGaSe}{GGS90}
\newcommand{\rAnHoGa}{AHG03}

\begin{document}
\title{Three-component modeling of C-rich AGB star winds}
\subtitle{III. Micro-physics of drift-dependent dust formation}
\author{C.~Sandin\inst{1,2}\and S.~H\"ofner\inst{1,3}}
\institute{Department of Astronomy and Space Physics,
Uppsala University, Box 515, S-751 20 Uppsala, Sweden \and
Astrophysikalisches Institut Potsdam, An der Sternwarte 16, D-14482 Potsdam, Germany \and
NORDITA, Blegdamsvej 17, DK-2100 Copenhagen \O, Denmark}
\offprints{C.~Sandin, \email{CSandin@aip.de}}
\date{Received April 2003/Accepted xxx}

\abstract{
A proper treatment of the non-equilibrium dust formation process is crucial in models of AGB star winds. In this paper the micro-physics of this process is treated in detail, with an emphasis on the effects of drift (drift models). We summarize the description of the dust formation process and make a few additions to previous work. A detailed study shows that different growth species dominate the grain growth rates at different drift velocities. The new models show that the net effect of drift is to significantly increase the amounts of dust, seemingly without affecting the mean wind properties, such as e.g., the mass loss rate. In some cases there is several times more dust in drift models, compared to the values in the corresponding non-drift models. We study the formation of a dust shell in the inner parts of the wind and find that drift plays an active role in accumulating dust to certain narrow regions. In view of the results presented here it is questionable if drift -- under the current assumptions -- can be ignored in the grain growth rates.
\keywords{hydrodynamics -- radiative transfer -- stars: AGB and post-AGB -- stars: mass-loss -- stars: variables:general}}

\authorrunning{C.~Sandin \& {S.~H\"ofner}}
\titlerunning{Three-component modeling of AGB star winds. III}
\maketitle

\section{Introduction}\label{sec:introduction}
A crucial aspect in wind models of asymptotic giant branch (AGB) stars is the description of dust formation. It is the radiation pressure on dust that with the support of the enclosed pulsating star is believed to form and drive the most massive winds in these stars. A detailed time-dependent treatment of the dust formation is necessary since it occurs far from equilibrium.

In the second article in this series, \citet[henceforth {\rSaHob}]{SaHo:03b}, we carried out a thorough study of the effects of grain drift on the average outflow properties of several types of time-dependent wind models. The results of wind models allowing drift (drift models) were compared with the respective non-drift models. A main finding was that drift, in most cases, modifies the wind structure to a significant degree concerning outflow properties and their temporal variability. In particular wind models that use a more realistic gas opacity are affected. The work presented in this article is based on the model description given in {\rSaHob}. Grain drift has so far been dynamically included through the use of a separate equation of motion for the dust. It was, however, not included in the processes describing dust formation. In regard of the drift induced changes found in previous results it is questionable if this treatment is adequate.

In this article we carry out a closer study of the micro-physics of dust formation in the wind forming region of time-dependent models, allowing drift. We begin by modifying the description of the grain growth process. Thereafter, results are discussed to assess the r\^ole of drift to the wind formation -- and the formation of dust shells. The purpose of this article is to focus on the understanding of the wind formation using a few typical models. A study closely related to this paper by \citet[henceforth {\rKrSe}]{KrSe:97} was concerned with the grain size distribution and dust formation in stationary models including drift; their conclusions are different form those found here. The detailed treatment of the dust material properties in time-dependent wind models was recently addressed in a study carried out by \citet[henceforth {\rAnHoGa}]{AnHoGa:03}. The results showed significant differences depending on the adopted properties of the dust. 

The modifications we carry out to include drift in dust formation processes are first described in Sect.~\ref{sec:physics}. Then the modeling procedure is presented together with a discussion on averaged outflow properties in Sect.~\ref{sec:results}. The consequences of allowing drift for the formation of a dust shell, and the details of the micro-physics of dust formation are discussed in Sect.~\ref{sec:discussion}; followed by the conclusions in Sect.~\ref{sec:conclusions}.

\section{Including drift in the dust formation description}
\label{sec:physics}
Like in the earlier articles in this series we distinguish between three interacting physical components of the wind. The components are the gas, the dust, and the radiation field. Each of these is described by coupled conservation equations that include exchange of mass, energy and momentum between all three components. A thorough description of the physical system, the gas-dust interaction, and the numerical method was given in {\rSaHoa}. The effects of stellar pulsations and an improved treatment of the gas opacity were added in {\rSaHob}. The work presented here is based on this most recent formulation. New in this article is the improved description of dust formation, which is extended to include effects of drift.

The dust component is assumed to consist of spherical particles made of amorphous carbon. Dust formation is described using the so-called moment method (\citealt{GaKeSe:84}, \citealt{GaSe:88}, and \citealt{GaGaSe:90} henceforth {\rGaGaSe}), involving four moment equations. The moments represent certain (average) properties of the grain size distribution function (see e.g.~{\rSaHob}, Table~1). Assumptions of this method are that grains are large enough that their thermodynamic properties do not depend on the grain size, and only molecules with a few monomers contribute significantly to the growth process. The moment equations can be written as (cf.\@ {\rGaGaSe}),
\begin{eqnarray}
\frac{\partial}{\partial t}K_0+\nabla\cdot(K_0v)&=&\mathcal{J}
\label{eq:phmom0}\\
\frac{\partial}{\partial t}K_n+\nabla\cdot(K_nv)&=&
 \frac{n}{d}\frac{1}{\tau}K_{n-1}+N_{\mathrm{l}}^{n/d}\mathcal{J}\::\:(1\le n\le d)\,.
\label{eq:phmom1}
\end{eqnarray}
Here $d$ is $3$ for spherical grains, $N_{\mathrm{l}}$ is the lower size-limit of macroscopic grains (we use $N_{\mathrm{l}}=1000$ carbon atoms), and $v$ is the dust velocity. Equation~(\ref{eq:phmom0}) describes the grain production, and the other three equations different properties of grains. In the form these equations are presented $\mathcal{J}$ is the grain nucleation rate. Nucleation describes the interchange of particles between the dust and gas phases (i.e.~the formation of seed particles). Like in previous work we adopt a stationary description ~\citep[cf.][Sect.~4]{GaSe:88}, using classical nucleation theory. We are, however, aware that this is a problematic point and that this assumption introduces quantitative uncertainties in the models. For a recent discussion on the nucleation rate and issues related to currently available descriptions see, e.g., {\rAnHoGa} and references therein. $1/\tau$ denotes the net grain growth rate (see next subsection).

Like in several of the earlier wind model articles we adopt a C-rich equilibrium chemistry. The atomic and molecular species involved in the dust formation are H, H$_2$, C, {\Ct}, {\CtH}, and {\CtHt}, where the last four contribute to grain formation processes. Dust formation has so far been treated accounting for nucleation, homogeneous growth, thermal evaporation, chemical growth, and chemical sputtering (see the following subsection). One assumption used in the description of these processes is that drift between gas and dust is negligible. In this study drift is the key feature and in the following subsections we discuss its implementation. In particular the net growth rate, $\tau^{-1}$, is affected.

\subsection{The treatment of grain growth neglecting drift}\label{sec:phPCgrgr}
In this subsection we describe grain growth without drift in order to identify the terms that need to be modified when allowing drift. {\rGaGaSe} wrote the grain growth rate in the following form,
\begin{eqnarray}
\left.\frac{1}{\tau}\right|_{\mathrm{PC}}&=&
  \frac{1}{\tau_{\mathrm{gr,h}}}
 -\frac{1}{\tau_{\mathrm{ev,h}}}
 +\frac{1}{\tau_{\mathrm{gr,c}}}
 -\frac{1}{\tau_{\mathrm{sp,c}}}\,.
\label{eq:phPCgrto}
\end{eqnarray}
The first term, $1/\tau_{\mathrm{gr,h}}$, describes homogeneous growth by addition of carbon $i$-mers (C$_i$; monomers). The second term, $1/\tau_{\mathrm{ev,h}}$, represents thermal evaporation from the surface of grains; as such it only depends on the properties of the dust. Reactions involving the two molecules {\CtH} and {\CtHt} are described by chemical growth (also called heterogeneous growth) $1/\tau_{\mathrm{gr,c}}$ and chemical sputtering $1/\tau_{\mathrm{sp,c}}$. The physical form of these terms is, when chemical equilibrium and a negligible drift are assumed, as follows,
\begin{eqnarray}
\frac{1}{\tau_{\mathrm{gr,h}}}
 &=&\sum_{i=1}^{I}iA_1\alpha_i\cdot \vtp_{i}f(i,t)\label{eq:phPCgrh}\\
\frac{1}{\tau_{\mathrm{ev,h}}}
 &=&\sum_{i=1}^{I}iA_1\alpha_i\cdot \vtp_{i}f(i,t)\cdot\frac{1}{S^i}
 \frac{\mathcal{K}_i(T_{\mathrm{d}})}{\mathcal{K}_i(\teg)}
 \sqrt{\frac{\teg}{T_{\mathrm{d}}}}\label{eq:phPCevh}\\
\frac{1}{\tau_{\mathrm{gr,c}}}
 &=&\sum_{i=1}^{I'}iA_1\sum_{m=1}^{M_i}
 \alpha_{i,m}^{\mathrm{c}}\cdot \vtp_{i,m}n_{i,m}\label{eq:phPCgrc}\\
\frac{1}{\tau_{\mathrm{sp,c}}}
 &=&\sum_{i=1}^{I'}iA_1\sum_{m=1}^{M_i}
 \alpha_{i,m}^{\mathrm{c}}\cdot \vtp_{i,m}n_{i,m}\cdot\nonumber\\
 &&\frac{1}{S^i}\frac{\mathcal{K}_{i,m}^{\,\mathrm{r}}(\teg)}{\mathcal{K}_{i,m}^{\,\mathrm{r}}(T_{\mathrm{d}})}\frac{\mathcal{K}_{i,m}(T_{\mathrm{d}})}{\mathcal{K}_{i,m}(\teg)}\,.\label{eq:phPCspc}
\label{eq:phPCgrgr}
\end{eqnarray}
The summation over $i$ labels the $I$ different carbon $i$-mers accounted for, in this case C and {\Ct}. Likewise $m$ labels the $M_i$ different reactions for molecules involving $I'$ carbon atoms, in this case {\CtH} and {\CtHt} for $i=2$. Moreover the gas and the dust temperatures are specified by {\teg} and $T_{\mathrm{d}}$. $\mathcal{K}_i$, $\mathcal{K}_{i,m}$, and $\mathcal{K}^{\,\mathrm{r}}_{i,m}$ are the dissociation constants that can be used to calculate partial pressures of the relevant molecules if chemical equilibrium is assumed (cf.\@ {\rGaGaSe} and references therein). The average sticking and reaction efficiencies are given by $\alpha_i$ and $\alpha_{i,m}^{\mathrm{c}}$, respectively (see Sect.~\ref{sec:phDrgrvs}). $S$ denotes the super-saturation ratio,
\begin{eqnarray}
S=\frac{P_{\mathrm{C}}(\teg)}{P_{\mathrm{C,sat}}(T_{\mathrm{d}})}
\end{eqnarray}
where $P_{\mathrm{C}}(\teg)$ is the actual partial pressure of carbon atoms in the gas phase. The denominator, $P_{\mathrm{C,sat}}(T_{\mathrm{d}})$, is the vapor (saturation) pressure of carbon atoms over a solid carbon surface, specified at the dust temperature $T_{\mathrm{d}}$. Like in earlier work we use \citep[Sect.~7]{GaSe:88},
\begin{eqnarray}
\log P_{\mathrm{C,sat}}=-\frac{86300}{T}+32.89
\end{eqnarray}
which is derived for graphite and is valid in the temperature regime $500\,\mbox{K}\lesssim T\lesssim1500\,\mbox{K}$. The (hypothetical) monomer surface area $A_1$ is given by,
\begin{eqnarray}
A_1=4\pi r_0^2=4\pi\left(\frac{3\mathcal{A}_{\mathrm{C}}m_{\mathrm{u}}}{4\pi\rho_{\mathrm{gr}}}\right)^{\frac{2}{3}}
\end{eqnarray}
where $r_0$ is the monomer radius, $\mathcal{A}_{\mathrm{C}}$ is the atomic weight of a carbon atom, $m_{\mathrm{u}}$ is the atomic mass constant, and $\rho_{\mathrm{gr}}$ the mass density of the condensed grain material. As in the previous articles we use the value for graphite, $\rho_{\mathrm{gr}}=2.25\,\mbox{g}\,\mbox{cm}^{-3}$. Note, however, that a value for amorphous carbon would be more consistent (cf.\@ {\rAnHoGa}). The number density of $i$-mers and the individual molecules involved in the building of dust grains are denoted by $f(i)$ and $n_{\mathrm{i,m}}$, respectively. The average thermal velocity of carbon $i$-mers, radicals, and molecules in the gas phase is given by the usual Maxwell-Boltzmann mean $\overline{v}_{i(,m)}$\footnote{The term $x_{i(,m)}$ is a short notation referring to both $x_i$ and $x_{i,m}$.} for each species. Particles that hit a grain surface cannot have a velocity directed such that they are moving away from the dust grain; it is easily seen that the integrated infalling flux of particles through a (planar) surface is given by $\Phi=f(i)\overline{v_i}/4$. For our purposes the velocity in Eqs.~(\ref{eq:phPCgrh})-(\ref{eq:phPCspc}) is defined as,
\begin{eqnarray}
\vtpr=\frac{\overline{v_{i(,m)}}}{4}=\frac{1}{4}\left(\frac{8k_{\mathrm{B}}\teg}{\pi \mathcal{A}_{i(,m)}m_{\mathrm{u}}}\right)^{\frac{1}{2}}
\label{eq:phPCvtp}
\end{eqnarray}
where $\mathcal{A}_{i(,m)}$ is the atomic weight of the particle species in question (see Table~\ref{tab:phDrgrgr}).

\subsection{The treatment of grain growth including drift}\label{sec:phDrgrgr}
The growth rates are modified in several ways when a relative motion between the gas and dust phases is allowed. Before we sum up the new rates in Sect.~\ref{sec:phDrgrtt} we discuss the necessary physical differences. Related works on stationary winds were carried out by \citet{DoSeGa:89} and {\rKrSe} (Sect.~2, who included fewer processes in the grain growth rates).

\subsubsection{The drift-dependent relative velocity and sticking coefficients}\label{sec:phDrgrvs}
The approximate relative velocity between particles moving with both a thermal velocity and a drifting velocity can be written as a root-mean-square of the sum of these two velocities~\citep[Sect.~5a]{Dr:80}. Since dust grains are much larger compared to gas particles, we again define our relative velocity by the same argument leading to Eq.~(\ref{eq:phPCvtp}),
\begin{eqnarray}
\wtpr=\frac{1}{4}\left(\overline{v_{i(,m)}}^2+\vdria^2\right)^{\frac{1}{2}}=\left(\vtpr^2+\frac{\vdria^2}{16}\right)^{\frac{1}{2}}
\label{eq:phDrvrtp}
\end{eqnarray}
where $\vdria=v(a)-u$ is the grain size dependent drift velocity, with $a$ denoting the grain radius. A drifting motion increases the flux of particles hitting a grain and thereby enhances the formation processes (see below).

\begin{figure}\centering
  \caption{The sticking coefficient $\alpha$ (Eq.(\ref{eq:phDrstick})) drawn as a function of the drift velocity {\vdri} for each of the four hydrocarbon species (upper panel). These species that are all involved in the grain growth rate chemistry are: C (dash-dot-dotted line), {\Ct} (dash-dotted line), {\CtH} (dotted line), and {\CtHt} (solid line). In addition the new/old ratio of the term $\wtp\alpha$ is shown in the lower panel. The figure shows that $\alpha$({\CtHt}) drops at a much lower drift velocity compared to the other three species, also note the steeply increasing grain formation efficiency (for all species) even at low drift velocities. The relative velocity {\wtp} is calculated using a gas temperature of $\teg=1200\,$K. Note, however, that {\wtp} is mostly dependent on the drift velocity when this is high.}\label{fig:phDrgrgr}
\end{figure}

Since we treat the drift velocity dependence of sticking coefficients ($\alpha_i$) and of reaction efficiencies ($\alpha_{i,m}^{\mathrm{c}}$) in the same way, we in the following refer to both these quantities as sticking coefficients, $\alpha_{i(,m)}$. Gaseous (hydro)carbon species may bond to radical surface sites on dust grains (chemisorption). The binding energy of the (hydro)carbon to the surface, $E_{\mathrm{b}}$, depends on the (unknown) surface morphology of the particle in question. The translational energy of the hitting particle must be adsorbed by the target, or it will bounce off. As a velocity dependent expression for the sticking coefficients we use the relation given by {\rKrSe} (note that the exponential form of this relation is based on a study involving the sticking of argon atoms onto argon-covered ruthenium),
\begin{eqnarray}
\alpha_{i(,m)}=\alpha_{\mathrm{C},i(,m)}\exp\left(-\left(\frac{0.5\mathcal{A}_{i(,m)}m_{\mathrm{u}}\wtpr^2}{4E_{\mathrm{b},i(,m)}}\right)^3\right)\label{eq:phDrstick}\,.
\end{eqnarray}
The nominator of the exponent in Eq.~(\ref{eq:phDrstick}) specifies the relative kinetic energy of the particle hitting a dust grain. In this expression it is assumed that a translational energy 4 times the binding energy can be adsorbed by a dust grain. We use binding energies based on quantum-mechanical calculations of the association of carbon species to sp$^3$ diamond surfaces. Binding energies and particle masses for each species are given in Table~\ref{tab:phDrgrgr}.

In contrast to the approach chosen by {\rKrSe} we adopt constant sticking coefficients $\alpha_{\mathrm{C},i(,m)}$ (smaller than one) in front of the exponential term in Eq.~(\ref{eq:phDrstick}). This is done to keep the same sticking coefficients as in previous models in the limit of zero drift velocities. For comparison reasons we adopt the same numbers we have used in earlier articles (e.g.\@, {\rSaHoa}, {\rSaHob}, and \citealt{HoFeDo:95} to mention a few), cf.\@ \citet{GaKeSe:84} and Table~\ref{tab:phDrgrgr} (these are the same sticking and reaction efficiencies used in Eqs.~(\ref{eq:phPCgrh})-(\ref{eq:phPCspc})). \citet{KrPaSe:96} and {\rKrSe} study sticking coefficients in more detail. The sticking coefficients as given by Eq.~(\ref{eq:phDrstick}) are plotted as a function of the drift velocity for each used hydrocarbon species in Fig.~\ref{fig:phDrgrgr}. Note that the sticking probability of {\CtHt} drops much faster with increasing drift velocity than those of the other species do; it quickly drops to zero when {\vdria} exceeds about 10\kms. The sticking probability of the other three species are more or less unaffected even when $\vdria=40\kms$.

The flux of gas particles hitting a grain surface increases with the drift velocity (Eq.~(\ref{eq:phDrvrtp})), thereby increasing the growth rate. If the drift velocity is too large, gas particles are too energetic to stick to a grain surface and instead bounce off, inhibiting further growth. However, the situation is ambiguous since different species contributing to the grain growth have different sticking probabilities. The increasing efficiency of grain growth/destruction with the drift velocity is illustrated using the term $\wtpr\alpha_{i(,m)}/\vtpr\alpha_{\mathrm{C},i(,m)}$ in Fig.~\ref{fig:phDrgrgr} (lower panel). Note that the grain growth efficiency increases fast with the drift velocity; it is about two times larger already for $\vdria\approx2\kms$ (more dust is indeed formed in the new drift models, Sect.~\ref{sec:resuopro}).

\begin{table}
\caption{Molecular properties of the (hydro)carbon species taking part in the grain growth process. From the left the columns give: the $i,m$ used in the growth rates; the molecule species; the atomic weight; the (constant) sticking coefficient (Eq.~(\ref{eq:phDrstick}); these are the same numbers used for $\alpha_i$ and $\alpha^{\mathrm{c}}_{i,m}$ in Eqs.~(\ref{eq:phPCgrh}-\ref{eq:phPCspc})); the binding energy for each species to a sp$^3$ carbon surface; and the corresponding source of binding energies in the literature.}\label{tab:phDrgrgr}
\begin{tabular}{llrrll}\hline\hline\\[-1.8ex]
$i,m$ & mol\@. & \multicolumn{1}{c}{$A$} & \multicolumn{1}{c}{$\alpha_{\mathrm{C}}$} & \multicolumn{1}{c}{$E_{\mathrm{b}}$} & references for $E_{\mathrm{b}}$\\
 & & \multicolumn{1}{c}{$[m_{\mathrm{u}}]$} & & \multicolumn{1}{c}{$[\mbox{eV}]$} \\[1.0ex]\hline\\[-1.8ex]
1   & C          & 12.011 & 0.37 & 4.0  & priv\@. comm\@. \;\dots\\
2   & {\Ct}      & 24.022 & 0.34 & 5.0  & \dots\; K\@. Larsson (2003)\\
2,1 & {\CtH}     & 25.030 & 0.34 & 6.5  & \citet{LaLuCa:93}\\
2,2 & {\CtHt}    & 26.038 & 0.34 & 0.27 & \citet{LaLuCa:93}\\[1.0ex]\hline
\end{tabular}
\end{table}

\subsubsection{Non-thermal sputtering; the erosion of dust grains by energetic abundant gas particles}\label{sec:phDrgrns}
A process unique for situations involving drift is non-thermal sputtering in which gas particles (projectiles) at high drift velocities are energetic enough to tear off carbon atoms from a grain surface (target). The corresponding rate of grain destruction is given by {\rKrSe} as,
\begin{eqnarray}
\frac{1}{\tau_{\mathrm{sp,n}}}=\pi r_0^2\vdria\sum_jn_jY_{\mathrm{sp},j}(E_j)
\label{eq:phDrspn}
\end{eqnarray}
where the sum over $j$ is taken for the different contributing particle species in the gas. The sputtering yield for each species is denoted by $Y_{\mathrm{sp},j}$. The number density and the kinetic energy of the projectile gas particles are given by $n_j$ and $E_j$, respectively. $E_j$ is simply,
\begin{eqnarray}
E_j=\frac{\mathcal{A}_jm_{\mathrm{u}}\vdria^2}{2}
\end{eqnarray}
where $\mathcal{A}_j$ is the atomic weight of the projectile particle. In the case of a solar chemical composition of the gas, contributions from other elements than hydrogen and helium can be neglected \citep[ see Fig.~1]{WoDoSe:93}. Furthermore, non-thermal sputtering is more or less negligible at drift velocities below $30\kms$. The drift velocity has earlier ({\rSaHoa},II) been found mostly to be lower than this. Nevertheless, non-thermal sputtering is included here since we want to get an idea of its relevance, if any, to the wind structure. For the yields in Eq.~(\ref{eq:phDrspn}) we adopt the empirical expressions given by~\citeauthor{BoRoBa:80} (\citeyear{BoRoBa:80}, \citeyear{BoRoBa:81}, also see \citealt{WoDoSe:93} and the additional references given therein),
\begin{eqnarray}
Y_{\mathrm{sp},j}=\left\{\begin{array}{l@{\::\:}l}0.0064\cdot\mathcal{A}_{\mathrm{C}}\gamma_j^{\frac{5}{3}}\left(\frac{\displaystyle E_j}{\displaystyle E_{\mathrm{th},j}}\right)^{\frac{1}{4}}\left(1-\frac{\displaystyle E_{\mathrm{th},j}}{\displaystyle E_j}\right)^{\frac{7}{2}}&E_j>E_{\mathrm{th},j}\\0&E_j\le E_{\mathrm{th},j}\end{array}\right.\label{eq:phDrspy}
\end{eqnarray}
Here $\gamma_j$ is given by,
\begin{eqnarray}
\gamma_j=\frac{4\mathcal{A}_{\mathrm{C}}\mathcal{A}_j}{(\mathcal{A}_{\mathrm{C}}+\mathcal{A}_j)^2}\,.
\end{eqnarray}
Moreover $\mathcal{A}_j$ is the atomic weight of the projectile particles (i.e.\@ about 1 and 4 for H and He, respectively). $E_{\mathrm{th},j}$ is the sputtering energy threshold and determines when sputtering starts to occur, it is specified as,
\begin{eqnarray}
E_{\mathrm{th},j}=\left\{\begin{array}{l@{\quad:\quad}l}\frac{\displaystyle E_{\mathrm{b,C}}}{\displaystyle \gamma_j(1-\gamma_j)}&\frac{\displaystyle\mathcal{A}_j}{\displaystyle\mathcal{A}_{\mathrm{C}}}\le0.3\\8E_{\mathrm{b,C}}\left(\frac{\displaystyle\mathcal{A}_j}{\displaystyle\mathcal{A}_{\mathrm{C}}}\right)^{\frac{2}{5}}&\frac{\displaystyle\mathcal{A}_j}{\displaystyle\mathcal{A}_{\mathrm{C}}}>0.3\end{array}\right.\,.
\end{eqnarray}
$E_{\mathrm{b,C}}(=4.0\,$eV) denotes the surface binding energy of individual carbon atoms on a dust grain. An alternative to Eq.~(\ref{eq:phDrspy}) is given in the semi-analytical yields presented by~\citet{TiMcKSeHo:94}.

\subsubsection{The growth rates accounting for drift}\label{sec:phDrgrtt}
All growth rates given in Eqs.~(\ref{eq:phPCgrh})-(\ref{eq:phPCspc}) describe the interaction of gas particles with dust grains. At the presence of a non-zero (average) drift velocity between these particles sticking coefficients and fluxes of particles hitting grain surfaces are modified. Thereby the homogeneous and chemical growth rates and the chemical sputtering are affected. The thermal evaporation (Eq.~(\ref{eq:phPCevh})) is, however, not affected as it only depends on the properties of the dust. Allowing drift we replace the total grain formation rate Eq.~(\ref{eq:phPCgrto}) with,
\begin{eqnarray}
\frac{1}{\tau}(a)=
  \frac{1}{\tau_{\mathrm{gr,h}}}
 -\frac{1}{\tau_{\mathrm{ev,h}}}
 +\frac{1}{\tau_{\mathrm{gr,c}}}
 -\frac{1}{\tau_{\mathrm{sp,c}}}
 -\frac{1}{\tau_{\mathrm{sp,n}}}=
 \frac{1}{\tau_{\mathrm{G}}}-\frac{1}{\tau_{\mathrm{sp,n}}}
\label{eq:phDrgrto}
\end{eqnarray}
where $\tau^{-1}_{\mathrm{G}}$ is the total drift-dependent growth rate excluding non-thermal sputtering. The growth rates given in Eqs.~(\ref{eq:phPCgrh})-(\ref{eq:phPCspc}) are replaced with,
\begin{eqnarray}
\frac{1}{\tau_{\mathrm{gr,h}}}
 &=&\sum_{i=1}^{I}iA_1\alpha_i(\wtp_{i})\cdot \wtp_{i}f(i,t)\label{eq:phDrgrh}\\
\frac{1}{\tau_{\mathrm{ev,h}}}
 &=&\sum_{i=1}^{I}iA_1\alpha_i\cdot \vtp_{i}f(i,t)\cdot\frac{1}{S^i}
 \frac{\mathcal{K}_i(T_{\mathrm{d}})}{\mathcal{K}_i(\teg)}
 \sqrt{\frac{\teg}{T_{\mathrm{d}}}}\label{eq:phDrevh}\\
\frac{1}{\tau_{\mathrm{gr,c}}}
 &=&\sum_{i=1}^{I'}iA_1\sum_{m=1}^{M_i}
 \alpha_{i,m}(\wtp_{i,m})\cdot \wtp_{i,m}n_{i,m}\label{eq:phDrgrc}\\
\frac{1}{\tau_{\mathrm{sp,c}}}
 &=&\sum_{i=1}^{I'}iA_1\sum_{m=1}^{M_i}
 \alpha_{i,m}(\wtp_{i,m})\cdot \wtp_{i,m}n_{i,m}\cdot\nonumber\\
 &&\frac{1}{S^i}\frac{\mathcal{K}_{i,m}^{\,\mathrm{r}}(\teg)}{\mathcal{K}_{i,m}^{\,\mathrm{r}}(T_{\mathrm{d}})}\frac{\mathcal{K}_{i,m}(T_{\mathrm{d}})}{\mathcal{K}_{i,m}(\teg)}\,.\label{eq:phDrspc}
\label{eq:phDrgrgr}
\end{eqnarray}

A fundamental difference, when compared to the earlier formulation, is that the rates now are grain size dependent through the drift velocity. When used with the moment equations (Eqs.~(\ref{eq:phmom0}) and (\ref{eq:phmom1})) this poses a problem since these do not readily allow for the use of grain size dependent rates. Moreover, our system of equations currently only includes one equation of motion for the dust, and thereby one mean grain velocity. A binned grain size distribution would require one additional equation of motion for each bin.

In {\rSaHoa} we argued that the dust velocity can be described using one mean quantity. With the current physical and methodological limitations we apply this assumption to grain growth as well, hence,
\begin{eqnarray}
\vdria=\vdri(\mdrad)=\vdri
\end{eqnarray}
where {\mdrad} is the average grain radius. Our study is hereby limited to qualitative effects of drift-dependent grain growth, and we do not study detailed effects associated with the grain size distribution.

\section{Modeling procedure and general results}\label{sec:results}

\subsection{Modeling procedure}\label{sec:resumpro}
The system of equations is discretized in the volume-integrated conservation form on a staggered mesh. The equations are solved implicitly using a Newton-Raphson algorithm where the Jacobian of the system is inverted by the Henyey method (cf.\@ {\rSaHoa}). In difference to the treatment of models calculated earlier we adopt a length scale of the artificial viscosity and artificial mass diffusion twice as large compared to the values used so far ($f=f_{\mathrm{d}}=7.0\times10^{-3}$, see Eqs.~(13) \& (14) in {\rSaHoa}). In some cases this higher value, in regions of very low dust densities, reduces large unwanted radial variations in the drift velocity (cf.\@ {\rSaHoa}, Sects.~3.2 \& 4.2); apart from that it has been found to only marginally affect the wind structure as a whole.

The modeling procedure is as follows. The wind model is started from a hydrostatic dust-free initial model where the outer boundary is located at about $2\,R_*$. All dust equations are switched on at the same time. Dust starts to form whereby an outward motion of the dust and the gas is initiated. The expansion is followed by the grid to about $25\,R_*$, where the outer boundary is fixed allowing outflow. The drift models evolve for about $50$-$200\,P$. The low density in the Planck mean models results in a `low mass' envelope that quickly is depleted of material. Therefore the time period for which the average outflow properties are calculated is shorter than the totally calculated time of the respective model (cf.\@ Sect.~4.2, {\rSaHob}). The instants studied in Sect.~\ref{sec:discussion} are selected at times before a significant fraction of the envelope is lost.

To describe the effects of stellar pulsations on the atmosphere we use a sinusoidal, radially varying inner boundary, located at about $0.91\,R_*$ (above the region where the $\kappa$-mechanism supposedly originates). An inflow of mass through the inner boundary is not permitted.

\subsection{Selection of model parameters}\label{sec:resmpar}
As was found in {\rSaHob}, wind models show a wide range of values in outflow properties and variability. And depending on what physical description is used for the gas opacity, the resulting density structure produces completely different winds. With this knowledge it is hard to choose a set of model parameters for one model that will be representative of a much larger set of wind models.

\begin{table}
\caption{Model parameters, cf.\@ Sect.~\ref{sec:resmpar}. The model names are given in Col.~1. Starting from the left the remaining columns specify: the stellar luminosity $L_*$; the effective temperature {\teff}; the pulsation period $P$; the pulsation amplitude {\deltaup}; and the carbon/oxygen ratio {\CtoO}. The stellar mass $M_*$ is set to $1.0\,M_{\sun}$ in all models.}
\label{tab:detmodel}
\begin{tabular}{lccccc}\hline\hline\\[-1.8ex]
model &$L_*$       &{\teff}     &$P$         &$\deltaup$&$\CtoO$ \\
      &$[L_{\sun}]$&$[\mbox{K}]$&$[\mbox{d}]$&$[\!\kms]$&\\[1.0ex]\hline\\[-1.8ex]
P13C16U6  & $1.3\times10^4$ & 2700 & 650 & 6 & 1.6\\
P13C14U6  & $1.3\times10^4$ & 2700 & 650 & 6 & 1.4\\[0.5ex]
P10C18U4  & $1.0\times10^4$ & 2790 & 525 & 4 & 1.8\\[1.0ex]\hline
\end{tabular}
\end{table}

Models calculated with a Planck mean gas opacity result in much more realistic density structures compared to the models calculated with a constant gas opacity (see e.g.\@ Sect.~2.2 in {\rSaHob}). The former treatment is adopted here. Moreover, in {\rSaHob} we found some models to be multi-periodic, meaning that the properties at the outer boundary vary with a period that is an integer multiple of the stellar piston period. Such a behavior is not seen in the typical model, and intentionally three sets of model parameters are selected that previously resulted in irregular wind variability; see Table~\ref{tab:detmodel}.

The winds of the two drift models P10C18U4 and P13C14U6 in {\rSaHob} were both found to give a smaller mass loss rate than that of P13C16U6. The latter model can be described as an `average' model in terms of other properties (see Tables~2 \& 5 and Sect.~4.2 in {\rSaHob} for further details). In the following model P13C16U6 is discussed in detail, and the two other models are presented for comparison with this model in Sect.~\ref{sec:resuopro}.

The modified physical descriptions of the grain growth rates discussed in Sect.~\ref{sec:physics} each affect the wind structure to different degrees. We want to assess the importance of each process by itself and therefore discuss four different versions of the model. One without drift -- a so-called position coupled (PC) model -- and three drift models. Of the latter three one is calculated without the modifications in the growth rates (i.e.\@ using Eq.~(\ref{eq:phPCgrto})). The second and third models adopt the drift modified growth rates, and the third in addition includes a description of non-thermal sputtering.

\begin{table*}
\caption{Quantities temporally averaged at the outer boundary for models P13C16U6, P13C14U4, and P10C18U4 (see Sect.~\ref{sec:results}). Physical features of the different versions of the model are indicated with the model name in the first column. Drift models are denoted by a '\emph{d}' and PC models by a '\emph{p}'. Moreover a '\emph{v}' indicates the use of the ``relative'' velocity given in Eq.~(\ref{eq:phDrvrtp}), and an '\emph{s}' the use of non-thermal sputtering. An unused feature is indicated by an '\_'. The following columns give the mean mass loss rate {\mmdot}, the mean terminal velocity {\muinf}, the mean degree of condensation {\mfcond}, the mean dust/gas density ratio {\mdrhog}, and the mean drift velocity {\mvdri}, respectively. In addition the standard deviation ({\sistd}) and the relative fluctuation amplitude $r\,(=\!\sistd/q)$ are specified for each quantity ($q$). The values shown in bold face of the quantities of the -\emph{dv\_} and -\emph{dvs} models indicate that they differ significantly (by $\ge10\%$) from the corresponding values of the -\emph{d\_\_} models. All models show an irregular temporal variability.}
\label{tab:resoprop}
\begin{tabular}{l@{\quad\quad\quad}r@{\ \ }rrcr@{\ \ }rrcr@{\ \ }rrcr@{\ \ }rrr}\hline\hline\\[-1.8ex]
   \multicolumn{1}{l}{model} & \multicolumn{2}{c}{$10^6$\,\mmdot} &&&
           \multicolumn{2}{c}{\muinf} &&&
           \multicolumn{2}{c}{\mfcond} &&&
           \multicolumn{2}{c}{\mdrhog} &&
           \multicolumn{1}{c}{\mvdri}\\
          & \multicolumn{2}{c}{$[M_{\sun}\,\mbox{yr}^{-1}]$} &&&
           \multicolumn{2}{c}{$[\kms]$} &&&
           \multicolumn{2}{c}{$[\%]$} &&&
           \multicolumn{2}{c}{$[10^{-4}]$} &&
           \multicolumn{1}{c}{$[\kms]$}\\
         &&\multicolumn{1}{c}{(\sistd)} & \multicolumn{1}{c}{{\it r}}&&
         & \multicolumn{1}{c}{(\sistd)} & \multicolumn{1}{c}{{\it r}}&&
         & \multicolumn{1}{c}{(\sistd)} & \multicolumn{1}{c}{{\it r}}&&
         & \multicolumn{1}{c}{(\sistd)} & \multicolumn{1}{c}{{\it r}}\\[1.0ex]\hline\\[-1.8ex]
P13C16U6-\emph{p\_\_} &3.9&(0.51)&{\it 0.13}&&14       &(0.23)&{\it 0.016}&&    21 &(9.6)&{\it 0.46}&&7.3     &(0.33)&{\it 0.045}&-\\
P13C16U6-\emph{d\_\_} &4.2&(3.0) &{\it 0.71}&&14       &(0.93)&{\it 0.066}&&    19 &(21) &{\it 1.1} &&9.3     &(23)  &{\it 2.5}&4.1\\[0.5ex]
P13C16U6-\emph{dv\_} &4.3&(3.2) &{\it 0.67}&&{\bf 19} &(1.3) &{\it 0.068}&&{\bf 33}&(32) &{\it 0.97}&&{\bf 26}&(93)  &{\it 3.5}&{\bf 4.0}\\
P13C16U6-\emph{dvs} &4.1&(3.4) &{\it 0.83}&&{\bf 19} &(1.4) &{\it 0.074}&&{\bf 32}&(35) &{\it 1.1} &&{\bf 26}&(82)&{\it 3.2}&{\bf 6.6}\\[2.0ex]
P13C14U6-\emph{p\_\_} &4.7&(2.3) &{\it 0.49}&&10       &(0.76)&{\it 0.076}&&     26 &(2.7)&{\it 0.10}&&    6.0 &(0.61)&{\it 0.10}&- \\
P13C14U6-\emph{d\_\_} &1.9&(2.0) &{\it 1.1} &&7.5      &(1.5) &{\it 0.20} &&     24 &(24) &{\it 1.0} &&    8.4 &(30)  &{\it 3.6} &10 \\[0.5ex]
P13C14U6-\emph{dvs}   &1.9&(1.5) &{\it 0.79}&&{\bf 8.4}&(3.4) &{\it 0.40} &&{\bf 39}&(25) &{\it 0.64}&&{\bf 11}&(19)  &{\it 1.7} &{\bf 8.4}\\[2.0ex]
P10C18U4-\emph{p\_\_} &1.1 &(0.49)&{\it 0.45}&&14&(0.77)      &{\it 0.055}&&16&(2.6)      &{\it 0.16}&&7.2&(1.2)     &{\it 0.17}&-\\ 
P10C18U4-\emph{d\_\_} &0.82&(0.12)&{\it 0.15}&&11&(0.24)      &{\it 0.022}&&12&(5.3)      &{\it 0.44}&&5.4&(2.9)     &{\it 0.54}&4.5\\[0.5ex]
P10C18U4-\emph{dvs}   &0.81&(0.32)&{\it 0.40}&&{\bf 19}&(0.50)&{\it 0.026}&&{\bf 46}&(21) &{\it 0.46}&&{\bf 21}&(14) &{\it 0.67}&{\bf 7.0}\\[1.0ex]\hline\\[-1.8ex]
\end{tabular}
\end{table*}

\subsection{Differences in averaged outflow properties caused by drift in dust formation}\label{sec:resuopro}
The main issue in {\rSaHob} was to study temporal variations of outflow properties which result from drift. To compare with those findings we here comment on outflow properties of three different kinds of drift models. The outflow properties ($q$), with the corresponding standard deviations ({\sistd}), and relative fluctuation amplitudes ($r=\sistd/q$), are given in Table~\ref{tab:resoprop}. The values of the models with the suffixes -\emph{p\_\_} and -\emph{d\_\_} are, with one exception, all identical to the corresponding PC model values given in {\rSaHob}, Table~5.  The exception is model P13C16U6-\emph{d\_\_} which has been recalculated here\footnote{The difference is due to the slightly different time-interval used in the calculation of the average properties, and the different length scale adopted for the artificial viscosity/diffusion.}.

The effects of drift-dependent dust formation excluding non-thermal sputtering are studied using model P13C16U6-\emph{dv\_}. The only apparent difference in the outcome of P13C16U6-\emph{dv\_} and P13C16U6-\emph{dvs} (which includes non-thermal sputtering) is the higher average drift velocity in the latter model. This higher value could be a result of how the mean of the drift velocity is calculated.  The higher value possibly arises due to the higher drift velocities found in the regions in front of shocks in model P13C16U6-\emph{dvs} (see Sect.~\ref{sec:discdusf}).

The values of the new sets of models, i.e.~those with a suffix -\emph{dv-} and -\emph{dvs}, differ significantly (defined as $\ge10\%$) in most quantities and fluctuation amplitudes from the corresponding values of the two old sets of models -\emph{d\_\_} and -\emph{p\_\_}. In particular, the higher values of both the degree of condensation and the dust/gas density ratio in all new winds indicate a more efficient dust formation when drift is accounted for, even at low drift velocities of only a few {\kms}. That the wind acceleration works more efficiently -- as a consequence of increased amounts of dust (seen in the dust/gas density ratio) -- is indicated in the higher terminal velocity of models P10C18U4-\emph{dvs}, P13C16U6-\emph{dv\_}, and P13C16U6-\emph{dvs}. The situation appears less certain in P13C14U6-\emph{dvs} where the value on the terminal velocity falls between the values of P13C14U6-\emph{p\_\_} and P13C14U6-\emph{d\_\_}.
As discussed in {\rSaHob}, PC model winds with a terminal velocity $\lesssim10\kms$ have no counterpart in drift models -- no wind is produced. Model P13C14U6-\emph{p\_\_} is a border line case, resulting in a much lower mass loss rate in P13C14U6-\emph{d\_\_}.
Due to the lower carbon abundance in this model (defined in the carbon/oxygen ratio), there is not enough material to form as much dust as in the other -\emph{dvs} and -\emph{dv\_} models. In contrast to the case of the terminal velocity the mass loss rate appears to be insensitive to the amounts of dust. All -\emph{dvs} (and -\emph{dv\_}) models show unchanged values when compared to the respective -\emph{d\_\_} models.

The issue of a larger variability in drift models was discussed in {\rSaHob} (Sect.~5.2). It was found that the variability in drift models mostly is larger than it is in PC models. Similarly the relative fluctuation amplitudes in the new -\emph{dvs} models in most quantities differ from the corresponding values of the -\emph{d\_\_} models. Concluding this section we note that the new results presented here seem to be difficult to reproduce without actually allowing drift in the calculation of dust formation.

\section{Discussion}\label{sec:discussion}
In this section we look closer at the region of wind formation. First we study the formation of a dust shell in detail in Sect.~\ref{sec:disckap}. Then, in Sect.~\ref{sec:discdusf}, we look closer at the effects on the micro-physics in dust formation caused by the addition of drift to the grain growth rates.

\begin{figure*}\centering
  \caption{Evolutionary sequence of the radial structure of the drift model P13C16U6-\emph{dvs} covering five instants of one complete dust formation cycle. The first column shows the dust formation at an arbitrarily selected time defined as $(t_0+)0.00\,P$ (in units of the stellar pulsation period). The three following columns are shown at $0.57\,P$, $1.12\,P$, and $1.63\,P$, respectively. The structure of the repeated dust formation cycle at $2.00\,P$ is shown by a solid line in the first column, illustrating the aperiodicity of dust formation in the current model. From the top the panels in each column show: {\bf a)} the gas velocity $u$; {\bf b)} the gas density $\rho$; {\bf c)} the gas temperature {\teg}; {\bf d)} the drift velocity {\vdri}; {\bf e)} the degree of condensation {\fcond}; {\bf f)} the dust/gas density ratio {\drhog}; and {\bf g)} the net growth rate $\tau^{-1}$. All panels are shown at the same scale in all four columns. The process illustrated in this figure shows that a nonzero drift velocity allows for the formation of narrow features in the dust already in the inner parts of the wind, cf.\@ Sect.~\ref{sec:disckap}.}
\label{fig:dusforma}
\end{figure*}

\subsection{The formation of a dust shell allowing drift}\label{sec:disckap}
Physical structures of drift model winds generally show a larger variability than corresponding PC model winds do (see Sect.~\ref{sec:resuopro}, and {\rSaHob}). This behavior is caused by an allowed dynamic accumulation of dust to narrower regions behind shocks; here we study how this accumulation might take place. The PC model P13C16U6-\emph{p\_\_}, for instance, shows a much smaller variability than the three drift models do and in fact reminds of a stationary wind. In the following we study the drift model P13C16U6-\emph{dvs}, which is calculated including both drift-dependent dust formation and non-thermal sputtering. As has been pointed out earlier (in Sect.~\ref{sec:results}) physical structures of different wind models show great variations and it is probably not safe to generalize details such as numbers and physical limits found for this wind model to all other possible cases. This study is a complement to the dust shell formation studies ignoring drift given in, e.g., \citet{FlGaSe:92,FlGaSe:95}.

One arbitrarily selected cycle of the dust shell formation is illustrated in Fig.~\ref{fig:dusforma}. The dust formation is in the following discussed in detail for each presented instant.

\emph{$1^{st}$ column, $0.00\,P$:} One of the shock waves (at $2.4\,R_*$) induced by the stellar pulsations has reached the region of efficient grain growth at $1.6-2.4\,R_*$, Figs.~\ref{fig:dusforma}a-b. The carbon in the gas quickly condenses into and onto grains, forming a new dust shell, Fig.~\ref{fig:dusforma}e; the current maximum degree of condensation, $\fcond\simeq20\%$, has been reached during $0.23\,P$ (not seen in the figure). Small amounts of dust grains are present inwards to a second shock at $1.6\,R_*$, inside of which it is too hot for grains to exist (and they instead evaporate). Grain growth is most efficient in the dense region behind the dust shell, where $r\lesssim2.4\,R_*$ (Fig.~\ref{fig:dusforma}g). The present dust heats the region behind the dust forming shell, causing a $100\,$K temperature step (backwarming), Fig.~\ref{fig:dusforma}c. The dust experiences an outwards directed radiative pressure originating in the central star, but the dust shell is not yet massive enough to push the gas outwards by itself. The drift velocity is low, Fig.~\ref{fig:dusforma}d, $\vdri\lesssim3\kms$ in all of the inner region; the relocation of dust is consequently slow as it takes the dust about $2.0\,P$ to travel a distance of $1.0\,R_*$ relative to the gas at this velocity.

\emph{$2^{nd}$ column, $0.57\,P$:} The dust shell (now at $3.3\,R_*$) has grown and is massive enough to drag the gas outwards as a consequence of the radiative pressure. A maximum of 80\% of the available carbon has currently condensed onto grains in the shell. Another dust shell is forming in the region behind the second shock at $2.0\,R_*$; a region of very efficient grain growth. The drift velocity is larger than in the previous column, reaching about $15\kms$. Note the depletion of dust in the forming dust shell behind $2.0\,R_*$ caused by a drift velocity reaching about $5.0\kms$.

\emph{$3^{rd}$ column, $1.12\,P$:} The dust in the previously forming dust shell (at $2.0\,R_*$ in the previous column) has been fully diffused into the region in front of the shock, now at $2.5\,R_*$. The remains are seen as a bump in the degree of condensation at about $4.0\,R_*$ (Fig.~\ref{fig:dusforma}e) behind the dust shell. The physical conditions, in the form of a low gas density and a small amount of dust grains, in the region between $2.5$ and $4.8\,R_*$ allow for a large drift velocity. Here it reaches values above $30\kms$, where non-thermal sputtering is active (see Sect.~\ref{sec:discdusf}). Relative to the gas the dust moves $2.3\,R_*$ in $0.5\,P$ at $30\kms$, and thereby allows for a quick accumulation to dense regions. This column shows the same instant that will be discussed in Sect.~\ref{sec:discdusf}.

\emph{$4^{th}$ column, $1.63\,P$:} The large drift velocity feature has moved outwards and almost caught up with the dust shell in front (now at $6.0\,R_*$). In the process, the dust in front of the feature has been ``swept up'' and is now mostly contained in a narrow shell which is merging with the original dust shell. The drift velocity behind the feature, at radii $<4.5\,R_*$, is low. A new dust shell is about to form at $2.3\,R_*$, behind a third gas shock (emitted $2.00\,P$ after the first shock discussed in the $1^{\mathrm{st}}$ column).

\emph{$2.00\,P$ (solid line in column 1):} In this frame two pulsation periods have passed since the filled line in the first column. The radial location of the next forming dust shell, now at $2.8\,R_*$, is different from the one forming at $0.00\,P$ (then at $2.4\,R_*$), indicating a non-periodic dust formation cycle.

The study of this dust shell formation cycle illustrates how dust may be accumulated to the regions behind shocks in drift models. While the wind is calculated out to $25\,R_*$, our discussion of the dust formation is limited to the innermost region of the wind, inwards of about $7\,R_*$. The outer region is subject to complex interactions of both physical and numerical character that complicates an interpretation. It is unclear if the conclusions drawn here can be extrapolated to a more extended region.

Next we discuss changes in the micro-physics, due to allowed grain drift in the dust formation, in more detail.

\begin{figure*}\centering
\caption{Illustration of the effects of drift in the inner parts of the stellar wind of model P13C16U6-\emph{d\_\_} (solid line) and P13C16U6-\emph{dvs} (dotted filled line). The panels show: {\bf a)} the gas velocity $u$; {\bf b)} the drift velocity {\vdri}; {\bf c)} the dust density {\rhod}; {\bf d)} the average grain radius {\mdrad}; {\bf e)} the gas temperature {\teg}; {\bf f)} the chemical growth (without sputtering) $\tau_{\mathrm{gr,c}}^{-1}(\mbox{C}_2\mbox{H})$ normalized to the net growth rate (excepting non-thermal sputtering) $\tau_{\mathrm{G}}^{-1}$; {\bf g)} the nucleation rate $J_*$; {\bf h)} the chemical growth (without sputtering) $\tau_{\mathrm{gr,c}}^{-1}(\mbox{C}_2\mbox{H}_2)$, also normalized (see {\bf f}); {\bf i)} the net growth rate excepting non-thermal sputtering $\tau_{\mathrm{G}}^{-1}$ (Eq.~(\ref{eq:phDrgrto})); and {\bf j)} the non-thermal sputtering $\tau^{-1}_{\mathrm{sp,n}}$ (Eq.~(\ref{eq:phDrspn})). All plots are drawn as a function of the stellar radius $R_*$ (lower axis), alternatively in astronomical units (upper axis). The dots on the contour represent individual grid points, a majority of which are located to the shocked regions. The gray horizontal lines are guides. Acetylene is normally the main growth species ({\bf h}) in the outer cool and dilute regions of the wind (modeled here). However, when the drift velocity is larger than about 10\kms ({\bf b}) the sticking coefficient $\alpha$ for acetylene quickly drops to zero (see Fig.~\ref{fig:phDrgrgr}), and the radical {\CtH} takes over this role ({\bf f}). Note that non-thermal sputtering (by helium; {\bf j}) is present when the drift velocity reaches about 30-35\kms.}
\label{fig:format}
\end{figure*}

\subsection{The effects of drift on the dust formation process}\label{sec:discdusf}
The physical conditions in cool C-rich stars are often suitable for efficient dust formation in a dynamic region around about $2.0$-$2.5\,R_*$. Different atomic and molecular species dominate the growth in different temperature (and pressure) regimes. At temperatures $\teg\lesssim1500\,$K it is {\CtHt}, for $1500\,\mbox{K}\lesssim\teg\lesssim1900\,$K {\CtH}, and for $\teg\gtrsim1900\,$K free C atoms dominate \citep[see][Fig.~1]{GaSe:88}. Of these the two former particles are the dominating carbon bearing species in the wind.

Figure~\ref{fig:format} shows the radial structure in the inner parts of model P13C16U6-\emph{dvs} at one instant of the dust formation cycle. This particular instant is selected as it simultaneously shows several distinguished features of the dust formation process (the same instant is shown in the third column in Fig.~\ref{fig:dusforma}).

A massive dust shell coinciding with a shock in the gas has formed and is moving away from the star at about $4.8\,R_*$ (Figs.~\ref{fig:format}a,c,d; also see the previous subsection). Due to the physical conditions, nucleation is currently sharply limited to the region between the two outermost shocks, where $2.4R_*\lesssim radius\lesssim4.8R_*$, and it is most efficient in the innermost (and most dense) part, Fig.~\ref{fig:format}g. The same region provides suitable conditions for a large drift velocity, which reaches 30\kms and above, by a low gas density (Fig.~\ref{fig:dusforma}b, Col.~3) and a small amount of dust grains, Fig.~\ref{fig:format}b. Few gridpoints are located in the region of a large drift velocity, resulting in a low resolution. The inner boundary of the peak in the drift velocity coincides with the shock front in the gas at $2.4\,R_*$.

The most abundant hydrocarbon molecule in most parts of the modeled envelope is {\CtHt} (acetylene), which also is the main growth species in most parts but the region around $2\,R_*$, where the radical {\CtH} instead dominates, Figs.~\ref{fig:format}f,h. Very small amounts of dust reside in the innermost region ($<1.5\,R_*$, Fig.~\ref{fig:format}c), where the chemistry and resulting properties therefore do not play a role. Contributions to the grain growth of the remaining two species, C and {\Ct}, are negligible in the current model.

Since the molecule {\CtHt} compared to, e.g., {\CtH} has a lower binding energy to the surface of a dust grain, it is in our calculations quickly exchanged as a primary growth element by other species when the drift velocity reaches $10\kms$ and above. In this case by {\CtH}, which sticks to the dust grains even if the drift velocity is larger than 40\kms (see Fig.~\ref{fig:phDrgrgr}). This is nicely illustrated in a comparison of the radial location of the peak of the drift velocity in Fig.~\ref{fig:format}b with the respective growth rates in Figs.~\ref{fig:format}f and h. Even though the growth rate of {\CtH} is about thirty times as efficient at $\vdri=30\kms$ its lower abundance results in a total growth rate $\tau^{-1}_{\mathrm{G}}$ lower by 0.5-1.0 orders of magnitude in the same region ($2.5R_*\lesssim\mbox{radius}\lesssim3.8R_*$), Fig.~\ref{fig:format}i. However, at high enough drift velocities non-thermal sputtering is active (see below) and the net growth rate $\tau^{-1}$ is negative, compare with the lower-most panel in Col.~3 Fig.~\ref{fig:dusforma}.

The two tall peaks in Fig.~\ref{fig:format}j, between $2.5$ and $3.5\,R_*$, are regions where the helium particles in the gas are energetic enough to erode dust grains. Non-thermal sputtering quickly becomes significant in comparison to the total growth rate $\tau^{-1}_{\mathrm{G}}$ when the drift velocity reaches values around 30-35\kms. As such this process is mostly present in regions of a low dust density where the drift velocity may be higher. And this is probably the reason to why the model properties (excepting the mean drift velocity itself) presented in Sect.~\ref{sec:resuopro} are found to be independent of it. A property that is affected is the average grain radius (Fig.~\ref{fig:format}d), which quickly decreases by two orders of magnitude and more in these regions.

That grain growth through the radical {\CtH} is not necessarily negligible is for example seen in a check of the relative growth rates of model P13C16U6-\emph{d\_\_}. In particular there seems to be a tendency towards a larger importance of {\CtH} in winds showing a large degree of variability in the structure. It is for example found to be less important in model P13C16U6-\emph{p\_\_}, while it is found do be the dominant growth species inwards of the innermost nucleation zone in model P13C16U6-\emph{d\_\_}; this is illustrated by the solid line in Figs.~\ref{fig:format}f-h. {\CtHt} is, however, still the dominant growth species in the improved drift models. Note that our use of an equilibrium gas chemistry may result in incorrect abundance ratios between different molecules. Reliable quantitative estimates of the relative contributions of {\CtH} and {\CtHt} to the growth rates can therefore not be made at present.

\section{Conclusions}\label{sec:conclusions}
An important part of recent models of AGB star winds is a time-dependent formulation that can account for, e.g., dust formation occurring in non-equilibrium, and formation and propagation of shock waves. To the authors' knowledge there has not been any time-dependent wind model that self-consistently treats grain drift in the dust formation process. This study is intended to fill this gap. We have covered both physical issues and numerical modeling, including a discussion on the implications and relevance for the wind structure. The new models are based on the wind model descriptions introduced by \citet[{\rSaHoa} \& {\rSaHob}, respectively]{SaHo:03,SaHo:03b}.

The new models presented here have been found to reproduce many properties of the previous drift models discussed in {\rSaHoa} and {\rSaHob}. New effects have, however, been introduced through the improved treatment. In particular the micro-physical details of the dust formation process are changed. An example is acetylene that normally is the main growth species in the wind~\citep[e.g.][]{GaSe:88}. This molecule is only weakly bound to the surface of dust grains and is therefore sensitive to the drift velocity. Its role in the grain growth process may be taken over by other more strongly bound growth species when the drift velocity increases above about 10\kms, partly preventing a decrease in the net growth rate. Dust destruction by non-thermal sputtering seems to be insignificant in view of the generally low drift velocities.

It should be noted that the drift-dependent dust formation in principle is grain size dependent through the drift velocity. The results that have been presented here are based on calculations using one mean drift velocity. A more consistent treatment might result in slightly different properties of the dust and the wind. Such a study, however, requires a significantly larger computational effort and is beyond the scope of the present paper.

The grain growth efficiency increases significantly with the drift velocity, even at values as low as a few {\kms}. Consequently the improved models produce significantly larger amounts of dust than the previous models did. However, it should be noted in this context that some of the micro-physical assumptions in the present models make detailed quantitative predictions of wind properties difficult. This concerns in particular the assumption of chemical equilibrium in the gas phase and the adopted sticking coefficients which may have a noticeable effect on the nucleation and growth rates. Further studies, beyond the scope of this paper, are needed to clarify these points. Still, in view of the results presented here, we conclude that the effects of drift in the grain growth rates cannot be simply ignored in wind models.

\begin{acknowledgements}
The calculations have been performed on the 12-processor HPV9000 placed at the Dept.\@ of Astronomy and Space Physics, financed through a donation by the \emph{Knut and Alice Wallenberg Foundation}. This work has been conducted within the framework of the research school on \emph{Advanced Instrumentation and Measurements} (AIM), at Uppsala University. AIM is financially supported by the \emph{Foundation for Strategic Research} (SSF).
\end{acknowledgements}

\bibliographystyle{aa}
\bibliography{CS_Refs}
\end{document}